\documentclass[
showpacs,
floatfix,
aps,prl,amsmath,
twocolumn,
groupaddress,
]{revtex4}

\usepackage{epsfig}
\usepackage{bm}
\usepackage{epsf}
\usepackage{array}

\usepackage[varg]{txfonts}

\newcommand{\be}{\begin{equation}}
\newcommand{\ee}{\end{equation}}

\newcommand{\bea}{\begin{eqnarray}}
\newcommand{\eea}{\end{eqnarray}}

\newcommand{\bes}{\begin{split}}
\newcommand{\ees}{\end{split}}

\newcommand{\req}[1]{Eq.~(\ref{#1})}

\newcommand{\rref}[1]{(\ref{#1})}

\newcommand{\kB}{k_{\rm B}}

\newcommand{\vare}{\varepsilon}
\newcommand{\mls}{\delta_1}
\newcommand{\esc}{\gamma_{\rm esc}}
\newcommand{\Nl}{N_{\rm l}}
\newcommand{\Nr}{N_{\rm r}}
\newcommand{\Nch}{N_{\rm ch}}

\newcommand{\tr}{{\rm tr}}

\newcommand{\Te}{T_{\rm e}}   %electron temperature
\newcommand{\Tf}{T_{\rm f}}   %EM temperature

\begin{document}
\title{Photovoltaic Current Response of Mesoscopic Conductors to
Quantized Cavity Modes}
\author{Maxim G. Vavilov}
\altaffiliation[Present address: ]{Department of Physics, University of
Wisconsin, Madison, WI 53706}
\affiliation{Department of Applied Physics, Yale University, New
Haven, CT 06520}
\author{A. Douglas Stone}
\affiliation{Department of Applied Physics, Yale University, New
Haven, CT 06520}

%\date{September 26, 2006}

\begin{abstract}

We extend the analysis of the effects of electromagnetic (EM) fields
on mesoscopic
conductors to include the effects of field
quantization, motivated by recent
experiments on circuit QED.
We  show that in general there is a photovoltaic (PV) current induced
by quantized cavity modes at zero bias
across the conductor.  This current depends on the average photon
occupation number and
vanishes identically when it is equal to the
average number of thermal electron-hole pairs.
We analyze in detail
the case of a chaotic quantum dot at temperature $\Te$ in contact
with a thermal EM field at temperature $\Tf$, calculating the RMS
size of the PV current as a function of
the temperature difference,
finding an effect $\sim \rm{pA}$.

\end{abstract}


\maketitle

Many quantum electronic devices for applications in metrology and quantum
information technology involve the interaction of electrons with
high frequency electromagnetic (EM) fields,
often the quantum devices act as detectors of this radiation~\cite{mestherm}.
In phase-coherent
(mesoscopic) devices there are quantum interference effects in
electron transport such
as the weak localization correction to the conductance and universal
conductance fluctuations which can in principle
be used to detect
radiation since it suppresses these effects~\cite{AAGS,AAKL,F87}.  In practice
the suppression of coherent transport by EM fields
is difficult to separate from the suppression by intrinsic
interactions due to the electron-electron and
electron-phonon couplings.

A more reliable mean of using mesoscopic conductors to detect EM
radiation is to look at the DC current induced by such a field at
zero voltage and temperature bias across the device, known as the
mesoscopic photovoltaic (PV) effect~\cite{FK,VAA,DCMH,VDCM}. This
effect arises in mesoscopic conductors because the phase-coherent
transmission through the device almost always violates parity
symmetry and the non-equilibrium distribution created by the EM
field sets up a steady-state current dictated by this parity
violation.  When the parity violation is due to random interference,
the sign of this current
will fluctuate from sample to sample and its root-mean-square (RMS)
size in this case depends on the power in the EM
field~\cite{FK,VAA,ZSA,Bpump,VDCM,PB}.  Hence after this PV current
is calibrated it can be used for detection of the power in the
incident EM field.

\begin{figure}
\epsfxsize=0.48\textwidth
%\vspace*{0.3\textwidth}
\centerline{\epsfbox{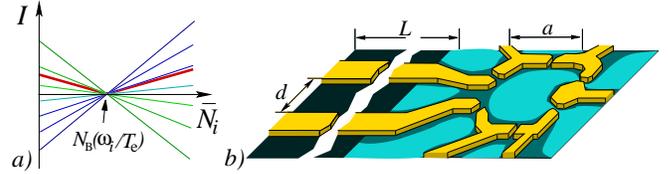}}
\caption{a) Schematic dependence of the PV current of a
mesoscopic conductor on the
average photon number $\bar N_i$ in a single mode resonator.
Narrow lines represent magnitude of the current for various
conductor realizations. The current vanishes for \emph{all}
realizations for $\bar N_i=N_{\rm B}(\omega_i/T)$.  Bold line is
the RMS value of the current over the ensemble, calculated for the
case of
a chaotic quantum dot. b) A possible experimental setup
for the observation of the
quantum photovoltaic effect. Two gates of a quantum dot
are terminals of a high quality microwave line of length $L$, that
form an EM resonator. To reduce dissipation,
the substrate of the resonator contains 2DEG (brighter area)
only in the vicinity of a quantum dot.
}
\label{fig1}
\end{figure}

The previous theoretical description~\cite{FK,VAA,ZSA,Bpump,PB,Kravtsov} of the
PV current has employed a classical
treatment of the EM fields, since this description was sufficient for
the systems studied experimentally~\cite{Kvon,linke,Switkes,DCMH,VDCM}. In this case
the RMS PV current is a monotonically increasing function of the EM
field power.
Recently a new
generation of electronic circuits was developed~\cite{cQED}, where a quantum
electronic device is coupled to an EM field of a high quality
electromagnetic resonator. If the resonator contains a small number
of photons  of the EM field and the lifetime of the photons is long, the
interaction of the EM field with electrons requires
a full quantum treatment, based on the laws of quantum electrodynamics,
leading to a new sub-field of quantum electronics known as
circuit QED.

In this Letter we investigate the properties of the mesoscopic
PV current that arises due to the electron interaction
with quantized EM fields.  We find that the net
current through the device can be represented as the sum of two
opposite contributions. One contribution is determined by
the average number $\bar N_i$ of photons with energy $\omega_i$
in modes $i$ of
the resonator. The second contribution is determined by
the number of thermal electron-hole pairs with energy
$\omega_i$ of resonator modes $i$, given by the
Bose distribution function $N_{\rm B}(\omega_i/\Te)$ at temperature $\Te$
of electrons in the leads; $N_{\rm B}(x)=1/(e^{|x|}-1)$.
We demonstrate that if both contributions
are taken into account,  the magnitude of the PV current, unlike the classical case,
is not
monotonic function of the strength of the EM field.  Instead
special conditions can be met when
these two contributions cancel each other and the PV
current vanishes for \emph{all} mesoscopic realizations of the device.
For the case of a thermal photon field the zero current state occurs
when the
temperature of the EM field and that of the electrons in the
leads are the same and
follows from the principle of detailed
balance.  For
an externally driven
single-mode cavity the zero current state also occurs
whenever the average number of photons $\bar N_i$ is equal to
the occupation number of
bosons $N_{\rm B}(\omega_i/\Te)$ at the electron temperature $\Te$,
independent of
the other properties of the full photon distribution in the
cavity.
In the classical description of EM fields only the first contribution is
found and the second contribution  due to electron-hole pairs
is missed.  As noted, in this case the PV current never vanishes
simultaneously for \emph{all} realizations and its RMS value
is a monotonic function of the power of the  EM
radiation~\cite{FK,SAA,VAA,PB}.
The non-monotonic behavior of the RMS value of the
PV current is an indication of the quantum behavior of
the EM fields.

A schematic depiction of the dependence of the PV current on the
number $\bar N_i$
of photons in a single mode resonator is shown in Fig.~\ref{fig1}~a,
for the case
of any mesoscopic conductor with fluctuating
transmission matrix.  For
a particular realization of the conductor, the current is a
linear function of $\bar N_i$ and changes its direction at
$\bar N_i=N_{\rm B}(\omega_i/T)$; at large values of $\bar N_i$ the dependence
on $\bar N_i$ will depart from linearity due to the suppression
of the coherence time by electron--photon scattering.
We note that the average value of the current with respect
to realizations of the mesoscopic conductor is zero, since the parity
violation of
transmission is zero on average.  In this case we
characterize the magnitude of
this current by the  RMS value of this current
averaged over the realizations, shown in Fig.~\ref{fig1}~a
by a bold line.

The generation of the photovoltaic current, studied here, is a
common phenomenon for out-of-equilibrium mesoscopic systems, which
function as ``quantum ratchets''. Other
similar phenomena are the Coulomb drag current \cite{NA} and the
current in mesoscopic metal rings coupled to out-of-equilibrium
electron~\cite{CK} or phonon~\cite{YK03} reservoirs.

\emph{Model ---}
The signatures of the quantum behavior of the EM field can be
observed in the PV current measurements for various
mesoscopic systems, such as quantum point contacts \cite{FK},
metal rings~\cite{Kravtsov}, metal wires or grains, and
semiconductor quantum dots~\cite{VAA}. The main requirement is
that the device has long coherence and inelastic relaxation times,
so that electron interference in propagation through the device
leads
to strong energy dependence and intrinsic parity violation in
transmission
to the left and right lead.  Any such mesoscopic device
will show
a PV current with the properties depicted in Fig.
~\ref{fig1}~a.

The specific case we will now treat in detail
is the PV current through
a semiconductor quantum dot with a few open channels
placed at the terminal of an electromagnetic resonator, see Fig.~\ref{fig1}~b.
The quantum dot is similar
to those used in charge pumping experiments~\cite{Switkes,DCMH}.
The electromagnetic resonator consists of
a microwave line of length $L$, characterized by high quality factor
and resonant
frequencies $\omega_i =i\pi c_*/L$~\cite{cQED}, where $c_*$
is the phase velocity of EM wave in the resonator and $i=1,2,\dots$.
The Hamiltonian of this system is
\be
\label{1}
{\cal H} ={\cal H}_{\rm d}
+{\cal H}_{\rm f} +{\cal H}_{\rm l}+{\cal H}_{\rm ld},
\ee
where $\hat {\cal H}_{\rm d}$ is the Hamiltonian
of the electrons in the dot
\be
{\cal H}_{\rm d}=\sum\limits_{n,m=1}^M
\psi^\dag_n  \left[ \hat H +
\sum_i\hat V_i
(a^\dag_i +a^{}_i)\right]_{nm}\!\!\!
\psi_m .
\label{Hd}
\ee
Here $\psi_n$ and $a_i $ are the annihilation operators
of electrons in the dot in state $n$ ($n=1,\dots, M$)
and photons in mode $i$
of the EM field, $M\times M$
Hermitian matrices $\hat H$  and $\hat V_i$
represent the stationary part of the electron Hamiltonian
and the electron coupling to mode $i$ of the EM field, respectively.
$\hat {\cal H}_{\rm f}$ describes the
evolution of the electro-magnetic field and can be written in terms
of photon annihilation and creation operators $a_i$ and $a^\dagger_i$
\be
\label{Hf}
\hat {\cal H}_{\rm f}=\sum_i \omega_i \left[a^\dagger_i
a^{}_i+1/2\right],
\ee
where $\omega_i$ is the energy of photon excitations.

The Hamiltonian
for electrons in the leads near the Fermi surface is
\be
\label{8}
\hat H_{\rm l}=v_{\rm F}\sum_{\alpha, k} k \psi^\dag_\alpha(k)\psi_\alpha(k),
\ee
where $\psi_\alpha(k)$ is the annihilation operator
of electrons in channel $\alpha$ of one of the leads.
The continuous variable  $k$ denotes electron momenta in the leads,
$v_{\rm F}=(2\pi \nu)^{-1}$ is the Fermi velocity,
and $\nu$ is the density of states per channel per spin at
the Fermi surface. In this Letter  we
consider the case when the voltage bias across the dot is zero.
The coupling of electron states in the dot  to states
in the leads  can be written as
\be
\label{7}
\hat H_{\rm ld}=\sum_{\alpha, n, k}\left( W_{n \alpha}\psi^\dag_\alpha
(k)
\psi_n+
{\rm H.c.}\right).
\ee
Here $\alpha$ labels channels in the leads, with
$1\leq \alpha\leq N_{\rm l}$ for the $N_{\rm l}$ channels in the left lead and
with $N_{\rm l}+1\leq \alpha \leq N_{\rm ch}$ for the $N_{\rm r}$ channels in
the right lead, $N_{\rm ch}=N_{\rm l}+N_{\rm r}$.
The coupling between electron states in the leads
and in the dot is described by $N_{\rm ch}\times M$
matrix $\hat W$.

\emph{Photovoltaic  current (PV) ---} We calculate the PV current
that flows through a quantum dot at zero temperature and voltage
biases. The interaction of electrons with the EM field results in
the deviation of the electron distribution function $n_{\rm
d}(\vare)$ in the dot from the Fermi distribution function $n_{\rm
F}(\vare)=[1+\exp(\vare/\Te)]^{-1}$ of electrons in the leads at
temperature $\Te$ and in a finite electric current through the
quantum dot. The direction and the magnitude of such current depend
on the mesoscopic violation of the left-right symmetry of the dot,
on the electron spectrum in the dot, and on the coupling strength of
electrons to the EM field. The derivation of the expression for the
current follows along the lines for the calculation of the current
through open quantum dots coupled to classical external
fields~\cite{VAA}. In the case of quantum fields,
the field acquires the off-diagonal matrix elements in
the Keldysh space, which can be easily taken into account within a
bilinear response.  As a result, we have
\be
\begin{split}
I= & \ e
\sum_i\sum_\pm \int
  J_i(\vare,\pm \omega_i)
  R_i(\vare,\pm \omega_i)d\vare
\end{split}
\label{I0}
\ee
The kernel $J_i(\vare,\omega)$ contains all the information about
electron motion in the dot
\be
J_i(\vare,\omega)=4\pi\nu\
\tr\!\! \left[
\hat W\hat\Lambda
\hat W^\dagger \hat G_{\rm r}(\vare)
\hat V_{i}{\rm Im}\{\hat G_{\rm r}(\vare-\omega)\}
\hat V_{i}
\hat G_{\rm a}(\vare)\right]
\label{Jvertex}
\ee
and corresponds to the triangle vertex diagram for the Coulomb drag~\cite{NA}, written
for the open dot geometry.
Function $\hat G_{\rm r}(\vare)$ is defined for a given
realization of $\hat H$ by
\be
\hat G_{\rm r}(\vare)=\frac{1}{\vare-\hat H-i\pi\nu\hat W\hat
W^\dagger};\quad
\hat G_{\rm a}(\vare)=[\hat G_{\rm r}(\vare)]^\dagger.
\label{eq:Gexact}
\ee
Here $\hat \Lambda=
(N_{\rm r}/N_{\rm ch}) \hat \Lambda_{\rm l}-
(N_{\rm l}/N_{\rm ch}) \hat \Lambda_{\rm r}$, where
$[\hat \Lambda_{\rm l}]_{\alpha\beta}=\delta_{\alpha\beta}$ for
$1\leq \alpha,\beta \leq N_{\rm l}$, and $[\hat \Lambda_{\rm
l}]_{\alpha\beta}=0$ otherwise;
$\hat \Lambda_{\rm r}=\hat 1-\hat \Lambda_{\rm l}$.
Equation~\rref{I0} takes into account spin degeneracy.

The function $R_i(\vare,\omega_i)$ is a combination of the
Fermi $n_{\rm F}$ and Bose $N_{\rm B}$ functions,
and the photon occupation number
$\bar N_i=\langle a^{\dagger}_ia^{}_i\rangle$
for the $i$ mode of the EM field:
\be
R_i(\vare,\omega)= 2\left[\bar N_{i} - N_{\rm B}(\omega /\Te)
\right]
\left[n_{\rm F}(\vare-\omega) -n_{\rm F}(\vare)
\right].
\label{eq:Rgeneral}
\ee
According to \req{eq:Rgeneral} the contribution to
the current $I$ from mode $i$ of the EM field
is linear in the average number $\bar N_i$
of photons with energy $\omega_i$ in mode $i$.
In particular, if only one mode of electromagnetic field
is coupled to electrons in the dot, the current $I$
can be used to determine the average number of photons in this
mode.

The structure of \req{eq:Rgeneral} can be understood from the
following schematic argument. The distribution
function in the dot $n_{\rm d}(\vare)$ is the solution of the
kinetic equation
\be
\frac{n_{\rm d}(\vare)-n_{\rm F}(\vare)}{\tau_{\rm esc}} =
\sum_i\left[ \Gamma_{i}^{\rm ab} (\vare) -
\Gamma_{i}^{\rm em} (\vare) \right].
\label{eq:kineq}
\ee
Here the left hand side describes the relaxation
of the distribution function $n_{\rm d}(\vare)$ due to electron
escape to the leads with characteristic escape time $\tau_{\rm esc}$
and the right hand side represents the imbalance between the rates
of absorption and emission of photons. These rates are determined
by $n_{\rm d}(\vare)$ and by the average number of photons $\bar N_i$
in mode $i$ of the electromagnetic field:
\be
\begin{split}
\Gamma_{i}^{\rm ab} & \propto  \bar N_i n_{\rm d}(\vare) ( 1- n_{\rm
d}(\vare+\omega_i)),\\
\Gamma_{i}^{\rm em} & \propto  [\bar N_i+1]  n_{\rm d}(\vare+\omega_i)
( 1- n_{\rm d}(\vare )).
\label{eq:rates}
\end{split}
\ee
Equations \rref{eq:kineq} and \rref{eq:rates} determine the
distribution function in the dot $n_{\rm d}(\vare)$.
To the lowest order in electron coupling to the EM field, we can
substitute $n_{\rm d}(\vare)=n_{\rm F}(\vare)$ in \req{eq:rates} and
obtain $\left[ \Gamma_{i}^{\rm ab} (\vare) -
\Gamma_{i}^{\rm em} (\vare) \right] \propto R_i(\vare,\omega_i)$.
A similar structure of the expression for the electric current was also
obtained in \cite{YK03}  for electron system coupled to an out-of-equilibrium
phonon reservoir.

When the thermal state of the EM field at temperature $\Tf$ is equal
to electron temperature $\Te$, $\bar N_i=N_{\rm B}(\omega_i/\Te)$,
the solution of \req{eq:kineq} is
$n_{\rm d}(\vare)\equiv n_{\rm F}(\vare)$. The latter equality is
the consequence of the detailed balance principle. Since the
distribution functions in the dot and in the leads are equal, the
PV current vanishes, as expected for a system in full
thermodynamic equilibrium.

\begin{figure}
\epsfxsize=0.3\textwidth
%\vspace*{0.3\textwidth}
\centerline{\epsfbox{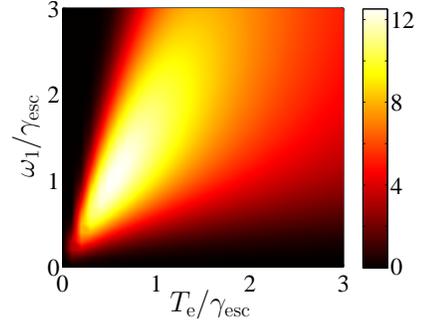}}
\caption{False color plot of the kernel ${\cal K}$ determining the
RMS PV current as a function of its arguments, the cavity mode frequency,
$\omega_1$, and the electron temperature, $\Te$, in units of
the electron
escape rate $\esc$ from the dot.  The maximum RMS current will occur at the
maximum of
${\cal K}$, which occurs when $\omega_1, \Te \sim \esc$.
}
\label{fig3}
\end{figure}

Note however that the PV current does \emph{not vanish} when there
are no  photons in the cavity, $\bar N_i = 0$. In this
case the current is driven by electron relaxation through spontaneous
emission of
photons  to the unoccupied modes of the EM field, and the
magnitude  of this current is determined by the matrix elements
$\hat V_i$ of zero-point fluctuations of the EM field.

Below we consider the case when the electromagnetic field
is in a thermal state at temperature $\Tf$ and
$\bar N_{i}=N_{\rm B}(\omega_i/\Tf)$.
Function $R_i(\vare,\omega)$ contains $N_{\rm B}(\omega_i/\Tf)-N_{\rm B}(\omega_i/\Te)$
and vanishes identically for $\Tf=\Te$.
%%%We can write
%%%\be
%%%R_{\rm th}(\vare,\omega)= \frac{\sinh[|\omega|(\Tf-\Te)/(2\Tf\Te)]}
%%%{2\cosh(\vare/2\Te) \cosh[(\vare-\omega)/2\Te]
%%%\sinh(\omega /2\Tf)}.
%%%\label{eq:Rth}
%%%\ee
For small deviations of
$\Tf$ from $\Te$, the current $I$
is linear in the temperatures difference:
$
I = B(\Tf-\Te)
$,
where the EM field thermopower coefficient $B$ is
\begin{eqnarray}
%\begin{split}
&&\!\!\!\!\! B=e \sum_i \int
  \left[J_i(\vare,\omega_i) r(\vare,\omega_i) -
  J_i(\vare,-\omega_i)r(\vare,-\omega_i)\right]
   d\vare ,\nonumber \\
&&\!\!\!\!\! r(\vare,\omega)= \frac{\omega/[4 \Te^2 \sinh(\omega/2\Te)]}
{\cosh(\vare/2\Te) \cosh[(\vare-\omega)/2\Te] }.
%\end{split}
\label{eq:B}
\end{eqnarray}

Equation \rref{eq:B} determines the value of the EM field
thermopower coefficient $B$ for a particular realization of the
Hamiltonian $\hat H$ of the quantum dot. Theoretical  and
experimental work~\cite{OSA,RMTex} has shown that lateral quantum dots
are well-described by a random-matrix (RM)
model of their transmission properties due to the
chaotic motion of electrons.
Therefore the theory of such systems has
focused on calculating averages of
relevant statistical quantities
over an appropriate RM ensemble.
Below we calculate the
statistical properties of $B$ with respect to a RM ensemble of $\hat H$.

\emph{Mesoscopic fluctuations of the current ---}
We calculate mesoscopic fluctuations of thermopower coefficient $B$
with respect to realizations of the $M\times M$ matrix $\hat H$
from a Gaussian ensemble of Hermitian matrices with $M\to\infty$,
characterized by the mean level spacing $\mls$:
$\langle H_{nm}H_{nm}^*\rangle=(M\mls^2/\pi^2)
\delta_{nn'}\delta_{mm'}
$ for a unitary  ensemble [The result is the same for
unitary and orthogonal ensembles].
The quantity of interest is the dimensionless quantity representing
the ensemble average product of
interaction matrices $\hat V_i$:
\be
\Gamma_{ij}=\frac{2\pi^2}{\Nch}\frac{\big{\langle}{\rm tr}\{\hat V_i\hat
V_j\}\big{\rangle}_{\rm ens}}{(M\mls)^2}
.
\label{eq:Gamma}
\ee
The coupling constants between electron states in the dot
and in the leads are
$W_{n\alpha}=\delta_{n\alpha}\sqrt{M\mls/\pi^2\nu}$ and result in
the following value of the electron escape rate from the dot:
$\esc=1/\tau_{\rm esc}=\Nch\mls/2\pi$.  Note that $\esc$ plays the
role for this
ballistic system of the Thouless energy of diffusive
systems, it sets the scale
of variation of the transmission with
electron energy.

The ensemble average value of $B$ is zero.
The variance of the PV current can be calculated
by diagrammatic RM technique~\cite{Vrev} and is
given by
\be
{\rm var}\ B  =  e^2\frac{\Nl\Nr}{\Nch^2}\sum_{ij}
\frac{\Gamma_{ii} \Gamma_{jj} }{\esc^2}
{\cal K}_{ij},
\label{eq:varB}
\ee
where the kernel ${\cal K}_{ij}$ is given by
\be
\begin{split}
{\cal K}_{ij}=&
K_{\omega_i,\omega_j}-K_{-\omega_i,\omega_j}
-K_{\omega_i,-\omega_j}+K_{-\omega_i,-\omega_j};
\\
K_{\omega_i,\omega_j} =&4\esc^2
\int
\left[
3+
\frac{2 [\Gamma_{ij}^2/(\Gamma_{ii}\Gamma_{jj})]\esc^2}
{\esc^2+(\vare -\vare'+\omega_j-\omega_i)^2}
\right]
\\
&\times
\frac{r(\vare ,\omega_i)r(\vare ,\omega_j)}
{\esc^2+(\vare -\vare')^2}d\vare  d\vare'.
\end{split}
\ee

To analyze the properties of the variance of the thermoelectric
coefficient $B$, we consider the case when only one mode $i=1$
of the electromagnetic field is coupled to electrons in the quantum dot.
We describe the properties of
${\cal K}_{11}={\cal K}(\omega_1/\esc, \Te/\esc)$, the contour
plot of this function is shown in Fig.~\ref{fig3}.

In the
low frequency limit of $\omega_1\ll \esc$, the power law
${\cal K}(\omega_1/\esc, \Te/\esc)\sim \omega_1^4$ is similar to the dependence
of the variance of the PV current induced by a
single-parameter classical perturbation~\cite{VAA}.
At low temperature $\Te\ll \omega_1$, the number of photons and
electron-hole pairs is exponentially suppressed and
${\cal K}\propto\exp(-\omega_1/\Te)$. At high
temperature $\Te\gg \esc$, the contribution to the thermoelectric
coefficient comes from electron states within thermal energies
and becomes self-averaged. As a result of such self-averaging, the
variance of $B$  decreases as ${\cal K}\sim 1/\Te$
as $\Te$ increases. To summarize,
${\cal K}(\omega_1/\esc, \Te/\esc)$ has a maximum at $T\propto
\omega_1$ at fixed $\omega_1$, see Fig.~\ref{fig3}.
The \emph{global} maximum of ${\cal K}(\omega_1/\esc, \Te/\esc)$
is ${\cal K}_{\rm max}\approx 12.5$ at
$\omega_1\approx 1.2 \esc$ and $\Te\approx 0.6 \esc$.
Thus, the largest effect will be observed when
$\Te$, $\omega_1$ and $\esc$ are all of the same order
of magnitude.

\emph{Conclusions ---}
We discuss experimentally achievable
values of the system parameters  (restoring $\hbar$ and $\kB$
in the equations below).
In experiments~\cite{cQED} $\omega_1/2\pi \sim 10$GHz
($\hbar\omega_1/\kB\approx 0.5$K) and
$\Tf\approx 30$mK. The escape rate $\esc$ for $\Nch\sim 1$ is
comparable with
$\mls/2\pi \hbar  \approx 2.5$GHz ($\mls/\kB\approx 0.12$K)~\cite{DCMH}.
%All of these energy scales are of the same
%order of magnitude and are optimal for the measurements
%of the EM thermopower effect.

To estimate the RMS value of the thermoelectric
coefficient $B_{\rm rms}=\sqrt{{\rm var} B}$, we write
$B_{\rm rms}\sim e\Gamma_{11}(\kB/\hbar)$, see \req{eq:varB},
where $\Gamma_{11} $ is defined by \req{eq:Gamma} and
can be expressed in terms of the magnitude
of zero-point electric field $E_1$ of the lowest
frequency mode $i=1$ of the EM resonator as
$\Gamma_{11}\approx e^2E_1^2a^2\tau_{\rm t}\tau_{\rm esc}/\hbar^2$~\cite{Vrev},
where $a$ is the diameter of the dot, see Fig.~\ref{fig1}, and
$\tau_{\rm t}=a/v_{\rm F}$ is the traversal time.
The zero-point electric field $E_1$ can be estimated
from the following equation $(E_1^2/4\pi)d^2 L\sim  \hbar\omega_1/2$,
where $L=\pi c_*/\omega_1$ is the resonator length, and gives
the estimate
$B_{\rm rms}\sim e\alpha_* (\kB/\hbar)(\omega_1^2\tau_{\rm t}\tau_{\rm esc})$,
where $\alpha_*=e^2/\hbar c_*$ and $a\sim d$. Compared to the
usual thermopower due to the temperature difference between the
leads~\cite{VS}, the EM thermopower at $\Te\simeq \hbar\esc/k_{\rm B}$
is suppressed by factor $\alpha_* \tau_{\rm t}\esc \ll 1$.
At $\omega_1/2\pi=10$GHz and $\tau_{\rm t}=4\cdot 10^{-12}$s, we have
$B_{\rm rms}\sim 35$pA/K, and for $|\Tf-\Te|\simeq 0.1$K the current
$\sim$pA is in the observable range.

We thank I. Aleiner, V. Manucharyan, and C. Marcus for discussions.
This work was supported by the W. M. Keck Foundation and
by NSF ITR DMR-0325580 and NSF Materials Theory grant DMR-0408638.

%\bibliography{ACtransport}

\begin{thebibliography}{10}

\bibitem{mestherm}
F. Giazotto \emph{et al.}, Rev.
  Mod. Phys. {\bf 78},  217  (2006).

\bibitem{AAGS}
B. Altshuler, A. Aronov, M. Gershenson, and Y. Sharvin, {\em Sov. Sci. Rev.
  A.:Phys.} (Harwood Academic Publishers, ADDRESS, 1987), Vol.~9, p.\ 223.

\bibitem{AAKL}
B.~L. Altshuler, A.~G. Aronov, D.~E. Khmelnitskii, and A.~I. Larkin, {\em
  Quantum Theory of Solids} (Mir publisher, Moscow, 1982).

\bibitem{F87}
V.~I. Falko, Zh. Eksp. Teor. Fiz. {\bf 92},  704  (1987).

\bibitem{FK}
V.~I. Fal'ko and D.~E. Khmelnitskii, Sov. Phys. JETP {\bf 68},  186  (1989).

\bibitem{VAA}
M.~G. Vavilov, V. Ambegaokar, and I.~L. Aleiner, Phys. Rev. B {\bf 63},  195313
   (2001).

\bibitem{DCMH}
L. DiCarlo, C.~M. Marcus, and J.~S. Harris, Phys. Rev. Lett. {\bf 91},  246804
  (2003).

\bibitem{VDCM}
M.~G. Vavilov, L. DiCarlo, and C.~M. Marcus, Phys. Rev. B {\bf 71},  241309
  (2005).

\bibitem{ZSA}
F. Zhou, B. Spivak, and B.~L. Altshuler, Phys. Rev. Lett. {\bf 82},  608
  (1999).

\bibitem{Bpump}
P.~W. Brouwer, Phys. Rev. B {\bf 58},  R10135  (1998).

\bibitem{PB}
M.~L. Polianski and P.~W. Brouwer, J. Phys. A {\bf 36},  3215  (2003).

\bibitem{Kravtsov}
V.~E. Kravtsov and V.~I. Yudson, Phys. Rev. Lett. {\bf 70},  210  (1993).

\bibitem{Kvon}
A.~A. Bykov, G.~M. Gusev, and Z.~D. Kvon, Sov. Phys. JETP {\bf 70},  742
  (1990).

\bibitem{linke}
H. Linke \emph{et al.},
  Europhys. Lett. {\bf 44},  341  (1998).

\bibitem{Switkes}
M. Switkes \emph{et al.}, Science {\bf 283},
  1907  (1999).

\bibitem{cQED}
A. Wallraff \emph{et al.}, Nature (London) {\bf 431},  162
  (2004).

\bibitem{SAA}
T.~A. Shutenko, I.~L. Aleiner, and B.~L. Altshuler, Phys. Rev. B {\bf 61},
  10366  (2000).

\bibitem{NA}
B.~N. Narozhny and I.~L. Aleiner, Phys. Rev. Lett. {\bf 84},  5383  (2000), the
  diagrammatic formalism for the Coulomb drag can be found in A. Kamenev and Y.
  Oreg, Phys. Rev. B \textbf{52}, 7516 (1995).

\bibitem{CK}
O.~L. Chalaev and V.~E. Kravtsov, Phis. Rev. Lett. {\bf 89},  176601  (2002).

\bibitem{YK03}
V.~I. Yudson and V.~E. Kravtsov, Phys. Rev. B {\bf 67},  155310  (2003).

\bibitem{OSA}
A. Altland, C.~R. Offer, and B.~D. Simons,  in {\em In proceedings of the 1997
  Summer School ``Disordered Systems and Quantum Chaos''}, edited by I.~V.
  Lerner (New York: Kluwer, ADDRESS, 1999).

\bibitem{RMTex}
I.~H. Chan \emph{et al.}, Phys.
  Rev. Lett. {\bf 74},  3876  (1995).

\bibitem{Vrev}
M.~G. Vavilov, J. Phys. A: Math. Gen. {\bf 38},  10587  (2005).

\bibitem{VS}
M.~G. Vavilov and A.~D. Stone, Phys. Rev. B {\bf 72},  205107  (2005).

\end{thebibliography}

\end{document}